\documentclass[12pt,preprint]{aastex}
\usepackage{amsmath}
\usepackage{graphicx}
\usepackage{color}
\usepackage{url}
\usepackage{CJKutf8}
\usepackage{ulem}

\slugcomment{Accepted by \aj}

\shorttitle{Hyperbolic Asteroids}
\shortauthors{Hui}

\begin{document}

\title{
Two Hyperbolic Baldheads in the Solar System: \\
2017 U7 and 2018 C2
}
\author{
\begin{CJK}{UTF8}{bsmi}
Man-To Hui (許文韜)$^{\dagger}$
\end{CJK}
}
\affil{
$^\dagger$Department of Earth, Planetary and Space Sciences,
UCLA, 595 Charles Young Drive East, Box 951567, 
Los Angeles, CA 90095-1567\\
}

\email{pachacoti@ucla.edu}

\begin{abstract}

We present a study of two newly discovered heliocentric hyperbolic objects -- 2017 U7 and 2018 C2. Both are possibly thermally evolved comets. Observations of the latter in 2018 March from Xingming Observatory revealed that it has a color similar to those of the long-period comets, Trojans, and D-type asteroids: $m_B - m_V = 0.75 \pm 0.03$, $m_V - m_R = 0.41 \pm 0.02$, and $m_R - m_I = 0.37 \pm 0.03$. A possible extremely faint coma of $\sim$9\arcsec~across was observed. The radial profile of 2018 C2 in comparison with those of the field stars helps confirm its cometary nature. Based on our convolution model, its mass-loss rate is estimated to be $0.7 \pm 0.2$ kg s$^{-1}$. Assuming geometric albedo $p_R = 0.04$, its effective radius is $4.4 \pm 0.5$ km, which means that the fraction of active area is merely $\sim$10$^{-5}$--10$^{-4}$. Our N-body dynamical simulations show that both objects are most likely dynamically old members from the Oort cloud. 2017 U7 has a $\sim$60\% chance to escape the solar system in the 1.5 Myr following its current perihelion passage due to a moderate close encounter with Jupiter in 2020 May. On the condition that no disintegration occurs, 2018 C2 will revisit the inner solar system 0.13-0.14 Myr later, with perihelion distance $\sim$2 AU.

\end{abstract}


\section{\uppercase{Introduction}}

Small bodies in the solar system are leftovers from planet formation, and are thought to be more primitive than the major bodies. Depending on where the small bodies used to reside before the gas had dissipated in the protoplanetary disc, compositional differences are expected. For instance, small bodies with long-term orbits within the ice line will likely have exhausted surface volatiles such as water ice, whereas objects that never entered the planetary region of the solar system may even retain supervolatiles such as CO and CO$_2$ to date due to the low temperature environment. However, it is likely that significant compositional reshuffling has occurred (Levison et al. 2009; Shannon et al. 2015), as dynamical studies suggest that the planets have undergone migration resulting from the gravitational pull of other growing planets and gas drag in the protoplanetary disc. During this process, insertion of primitive outer solar system objects into the asteroid belt may have taken place, and likewise, a portion of inner solar system materials may also have been scattered outward. While some became interstellar objects like 1I/2017 U1 (`Oumuamua), which was likely ejected from an extrasolar system (de la Fuente Marcos \& de la Fuente Marcos 2017), and thus may indicate that similar processes once occurred in our own system, the remaining material ended up constituting a spherical structure at the edge of the solar system -- the Oort cloud under influences from the galactic tides and random passing stars (Oort 1950). Although different dynamical models (e.g., Tsiganis et al. 2005; Walsh et al. 2011) are largely capable of reproducing the modern solar system architecture, discrepancies do exist, in terms of the capability of scattering inner solar system materials to the edge by migrating planets. Therefore, it is scientifically important to study the dynamical histories and compositions of the small bodies.

Recently, discoveries of two small bodies having heliocentric hyperbolic orbits -- 2017 U7 and 2018 C2 (hereafter ``17U7" and ``18C2", respectively) were announced by the Minor Planet Center (MPC).\footnote{The former was prefixed by ``A/" by the MPC, indicating an asteroidal appearance. The latter used to be prefixed by ``A/" too, yet was changed to ``C/" when this paper was in preparation. To avoid potential future changes, we are not using the prefixes here.} 17U7 was discovered by the Panoramic Survey Telescope and Rapid Response System (Pan-STARRS) on UT 2017 October 29.4 at mag $\sim$20. Subsequently, prediscovery observations from the same survey from 2017 August were identified. 18C2 was discovered by the Mt. Lemmon Survey, part of the Catalina Sky Survey, on UT 2018 February 05 at mag $\sim$20. Similarly, prediscovery images from the same survey were found, which helped extend the arc by roughly a week. The majority of the small bodies previously known to be in heliocentric hyperbolic orbits are comets having prominent activity. In contrast, 17U7 and 18C2 are apparently far less active. Despite that their heliocentric orbits are hyperbolic, the eccentricities are barely $e > 1$, suggesting that they are unlikely recent interstellar interlopers like `Oumuamua. If so, 17U7 and 18C2 are the first couple members of such population ever discovered. We thus feel the necessity to perform a study about them for a better understanding of objects of this type. 

Our paper is organized in the following order. Section \ref{sec_obs} describes observations and photometry analysis of 18C2. Orbit determination for 17U7 and 18C2 is presented in Section \ref{sec_orb}, and the study of their dynamical evolution is detailed in Section \ref{sec_dyn}. Discussions are held in Section \ref{sec_disc}, while Section \ref{sec_sum} presents a summary.

\section{\uppercase{Observations}}
\label{sec_obs}

Unfortunately, object 17U7 is currently at a small elongation from the Sun ($\lesssim 40$\degr), and will not be able to be observed in the following few months. So no physical observations of it can be gleaned. Nor did we find any positive serendipitous prediscovery archival observations using the Solar System Object Image Search (SSOIS; Gwyn et al. 2012) of the Canadian Astronomical Data Centre (CADC). But for 18C2, we were able to obtain Johnson system {\it B}, {\it V}, {\it R} and {\it I}-band images from Xingming Observatory, Xinjiang, China. These observations were taken through the Ningbo Bureau of Education and Xinjiang Observatory Telescope (NEXT), a 0.6-m f/8 Ritchey-Chr{\'e}tien telescope equipped with a 2k $\times$ 2k CCD. The exposure time was chosen to maximize the signal-to-noise ratio (SNR) of the target in each image while it would not be trailed by more than one image pixel, because the mount was tracking in a sidereal rate. The field-of-view of the images is $21\arcmin.5 \times 21\arcmin.7$, with an angular resolution of 0\arcsec.63 pixel$^{-1}$ in the bin $1 \times 1$ mode. Full width at half maximum (FWHM) of field stars in the images varied between $\sim$2\arcsec.0 and 2\arcsec.9. The observing geometry of 18C2 is summarized in Table \ref{tab_geo}.

\begin{figure}
\epsscale{0.8}
\begin{center}
\plotone{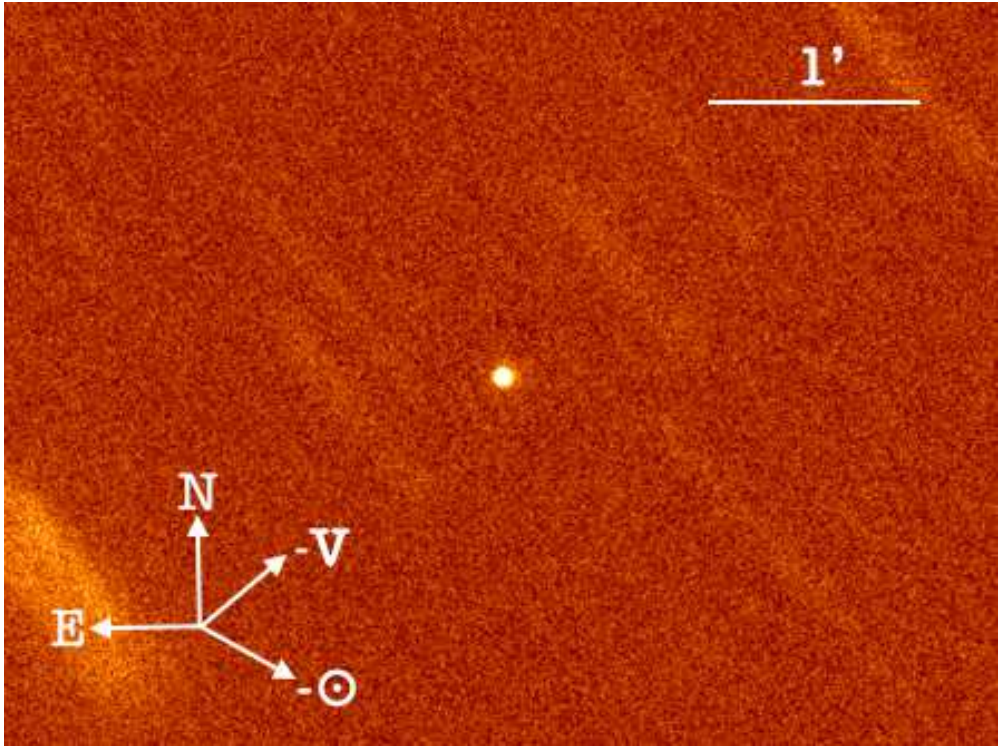}
\caption{
A median coadded {\it V}-band image of 2018 C2 from 25 individual images taken on UT 2018 March 19 by NEXT, Xingming Observatory, showing that the object has an extremely faint coma almost in the same level as the background noise. Indicated by a compass in the lower left corner, equatorial north is up and east is left. Also shown are the antisolar direction ($-\odot$) and the negative heliocentric velocity vector projected onto the sky plane ($-{\bf V}$). A scale bar is given at the upper right corner. The cropped image has a dimension of 4\arcmin.7 $\times$ 3\arcmin.5. The faint streaks are trails of background stars and galaxies.
\label{fig_18C2_180319V}
} 
\end{center} 
\end{figure}

The images were calibrated with bias, dark frames, and sky flat images in the corresponding bands. Uncleaned hot pixels together with cosmic rays, if any, were removed using the package LA Cosmic (van Dokkum 2001). Additional steps were taken for {\it I}-band images, as obvious fringes were seen. A minimum background was computed from the {\it I}-band images from the different nights with individual median background values offset to zero, and then was subtracted from the calibrated images. The resulting images have the fringe patterns largely removed. Finally, images from the same observing night in the same bandpasses were median coadded with registration on 18C2 to improve the SNR, and also on field stars. A possible extremely faint coma, which can be easily confused as background noise, of $\sim$9\arcsec~across was observed for 18C2 in the coadded images (see Figure \ref{fig_18C2_180319V} as an example).

\subsection{Photometry}
\label{sec_phot}

We used an aperture of 9 pixels (5\arcsec.7) radius to measure fluxes of the field stars. The sky background was measured using an annulus adjacent to the aperture, with inner and outer radii 8\arcsec.5 and 14\arcsec.2, respectively. The Sloan Digital Sky Survey (SDSS) Data Release 12 (DR12; Alam et al. 2015) was used for photometric calibration. As catalogued comparison stars are in the Sloan system, we first transformed the Sloan-system magnitudes to the corresponding Johnson system by means of relationships derived by Chonis \& Gaskell (2008). The zero-point was then solved on each coadded image with alignment on field stars. Finally, aperture photometry measurements of 18C2 on the coadded images with registration on its non-sidereal motion with the same photometry configuration. 

We did attempt to perform photometry of 18C2 in individual images, aiming to obtain its spin period. However, no statistically obvious periodic trend has been identified by the phase dispersion minimization (PDM; Stellingwerf 1978) algorithm, with the PDM parameter $\Theta \ge 0.87$. 

Table \ref{tab_phot} summarizes the photometry results. The mean color of 18C2 has $m_B - m_V = 0.75 \pm 0.03$, $m_V - m_R = 0.41 \pm 0.02$, and $m_R - m_I = 0.37 \pm 0.03$, where the uncertainties are propagated from errors of individual measurements. Thus, its color is only slightly redder than that of the Sun: $\left(m_B - m_V \right)_{\odot} = 0.65 \pm 0.01$, $\left(m_V - m_R \right)_{\odot} = 0.36 \pm 0.01$, and $\left(m_R - m_I \right)_{\odot} = 0.35 \pm 0.01$ (Ram{\'{\i}}rez et al. 2012), and similar to those of the long-period comets, D-type asteroids, and Trojans, yet significantly bluer than ultrared trans-neptunian objects (Jewitt 2015).

Absolute magnitudes of 18C2 in bandpass $\lambda$, denoted as $m_{\lambda} \left(1,1,0 \right)$, were computed from the corresponding apparent magnitudes, $m_{\lambda} \left(r_{\rm H}, {\it \Delta}, \alpha \right)$ by
\begin{equation}
m \left(1, 1, 0\right) = m\left(r_{\rm H}, {\it \Delta}, \alpha \right) - 5 \log \left(r_{\rm H} {\it \Delta}\right) + 2.5 \log \Phi \left(\alpha\right)
\label{eq_Hmag},
\end{equation}
\noindent where $r_{\rm H}$ is the heliocentric distance, ${\it \Delta}$ is the object-observer distance, both in AU, and $\Phi$ is the compound phase function normalized at phase angle $\alpha = 0\degr$, to which is contributed by the phase function of the nucleus $\Phi_{\rm N}$, and that of its dust coma $\Phi_{\rm C}$ (see Appendix \ref{apndx}):
\begin{equation}
\Phi \left( \alpha \right) = \frac{\Phi_{\rm N}\left( \alpha \right) + \eta_0 \Phi_{\rm C} \left( \alpha \right) }{1 + \eta_0}
\label{eq_phi}.
\end{equation}
\noindent Here, $\eta_0 \ge 0$ is the flux ratio between the coma and the nucleus observed at $\alpha = 0\degr$. We approximated $\Phi_{\rm N}$ by the HG formalism (Bowell et al. 1989) with slope parameter $G=0.15$, and $\Phi_{\rm C}$ by an empirical relationship by Schleicher (2010; \url{asteroid.lowell.edu/comet/dustphase.html}). Comparison between the surface profile of 18C2 and the field stars yielded $\eta_0 \approx 0.3$ from the observations (see Section \ref{sec_act}), with which the absolute magnitudes were obtained (Table \ref{tab_phot}). 

\begin{figure}
\epsscale{1.0}
\begin{center}
\plotone{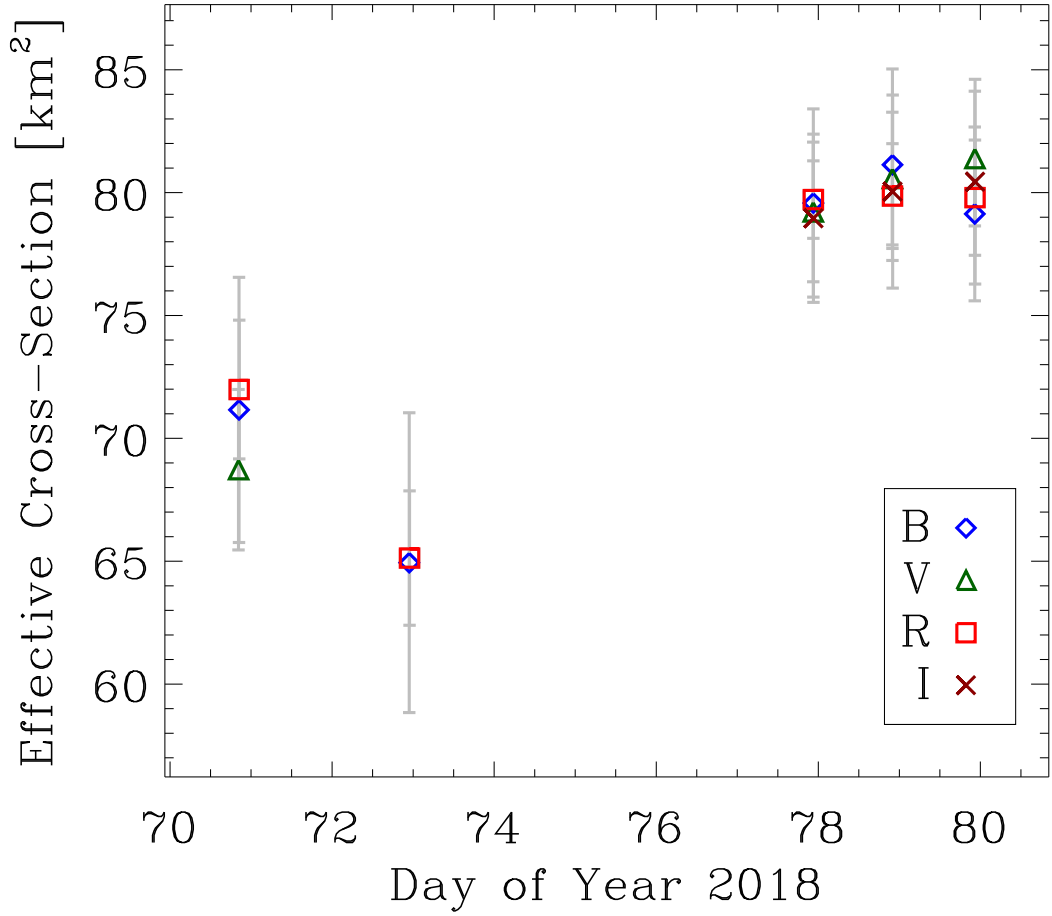}
\caption{
The effective cross-section of 2018 C2 as a function of time, with an assumed geometric albedo $p_{R} = 0.04$. Albedos in different bandpasses are scaled based on the mean color of the object. The error bars are propagated from the $1\sigma$ magnitude uncertainties. At a $3\sigma$ level, there is no compelling evidence that the effective cross-section is varying temporally.
\label{fig_18C2_t_vs_XS}
} 
\end{center} 
\end{figure}

The effective total cross-section of 18C2, $C_{\rm e}$, is related to the absolute magnitude by
\begin{equation}
C_{\rm e} = \frac{\pi r_{\oplus}^{2}}{p_{\lambda}} 10^{-0.4 \left[m_{\lambda} \left(1,1,0\right) - m_{\odot, \lambda}\right]}
\label{eq_xs},
\end{equation}
\noindent where $r_{\oplus} \approx 1.5 \times 10^{8}$ km is the mean heliocentric distance of the Earth, $p_{\lambda}$ is the geometric albedo, and $m_{\odot}$ is the apparent magnitude of the Sun at $r_{\oplus}$. Absolute magnitudes were all transformed to {\it R}-band with the derived mean color indices. The Sun has $m_{\odot,R} = -27.09$ (Williams 2018, \url{nssdc.gsfc.nasa.gov/planetary/factsheet/sunfact.html}). Assuming a typically low cometary albedo $p_R = 0.04$, the temporal variation of the effective total cross-section is shown in Figure \ref{fig_18C2_t_vs_XS}. The effective cross-section of the object appears to change versus time, however, taken into consideration the uncertainty, it is not statistically important. We estimated its nucleus radius to be $R_{\rm N} = \sqrt{C_{\rm e} / \left[ \pi \left(1 + \eta_0 \right) \right]} = 4.4 \pm 0.5$ km, where the uncertainty includes errors in magnitude data, the phase function, and $\eta$. An additional systematic bias may exist due to the assumed geometric albedo value, which has not been considered here. However, we can still confidently conclude that the nucleus of 18C2 is a few kilometres in radius.

\subsection{Activity}
\label{sec_act}

\begin{figure}
  \centering
  \begin{tabular}[b]{@{}p{0.3\textwidth}@{}}
    \centering\includegraphics[width=0.36\textwidth]{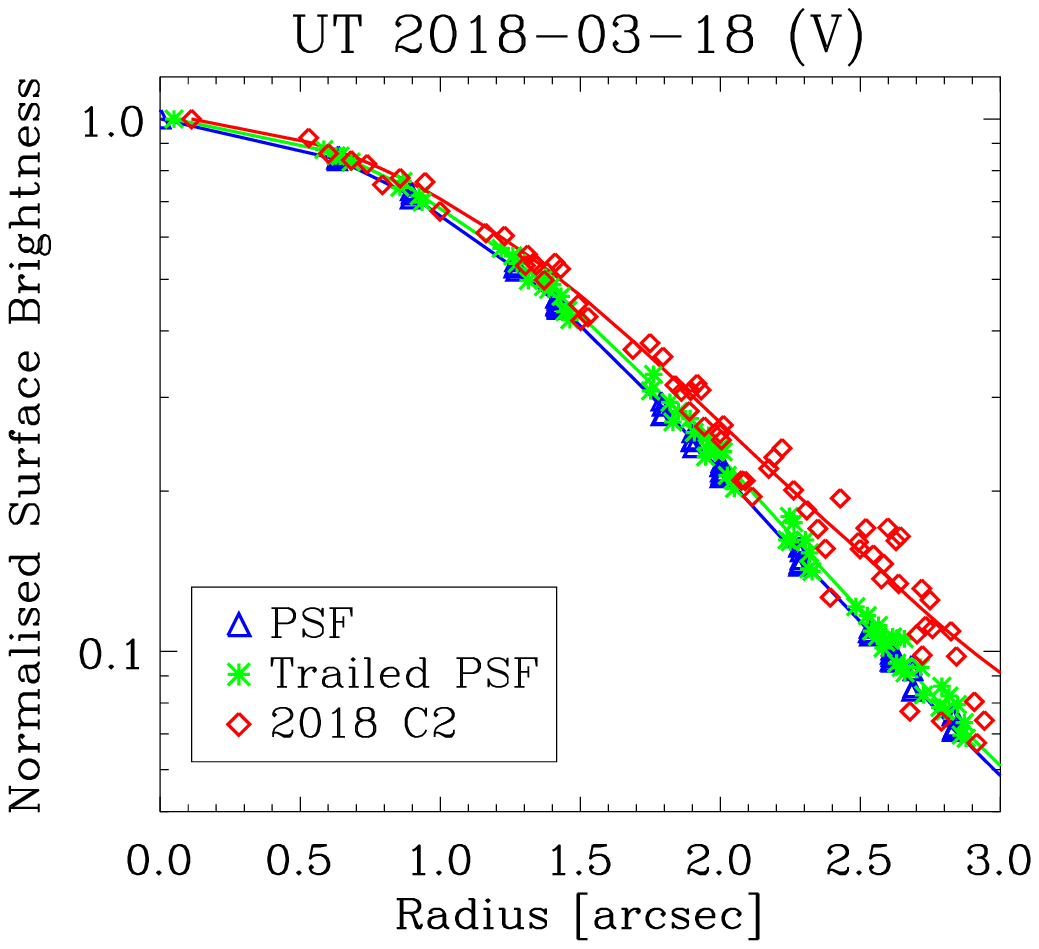} \\
    \centering\small ~~~(a)
  \end{tabular}%
  \quad
  \begin{tabular}[b]{@{}p{0.3\textwidth}@{}}
    \centering\includegraphics[width=0.36\textwidth]{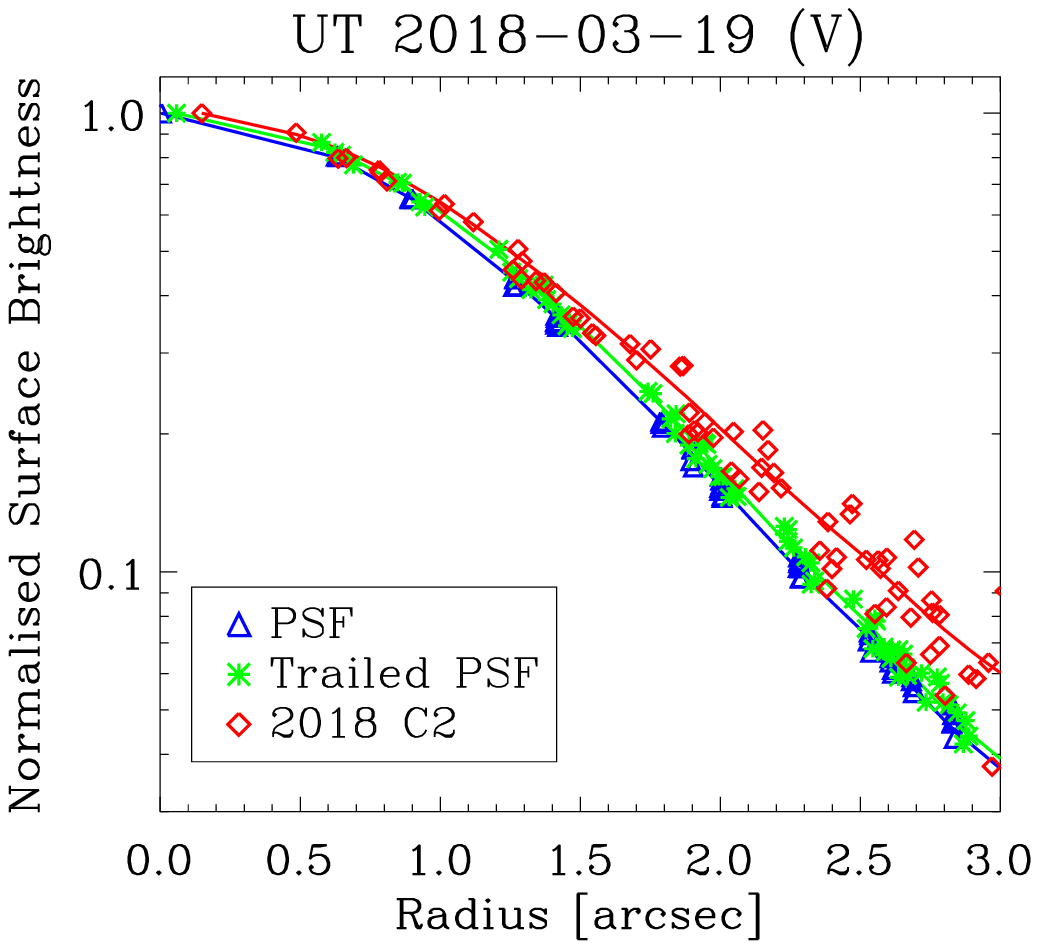} \\
    \centering\small ~~~(b)
  \end{tabular}
  \quad
  \begin{tabular}[b]{@{}p{0.34\textwidth}@{}}
    \centering\includegraphics[width=0.36\textwidth]{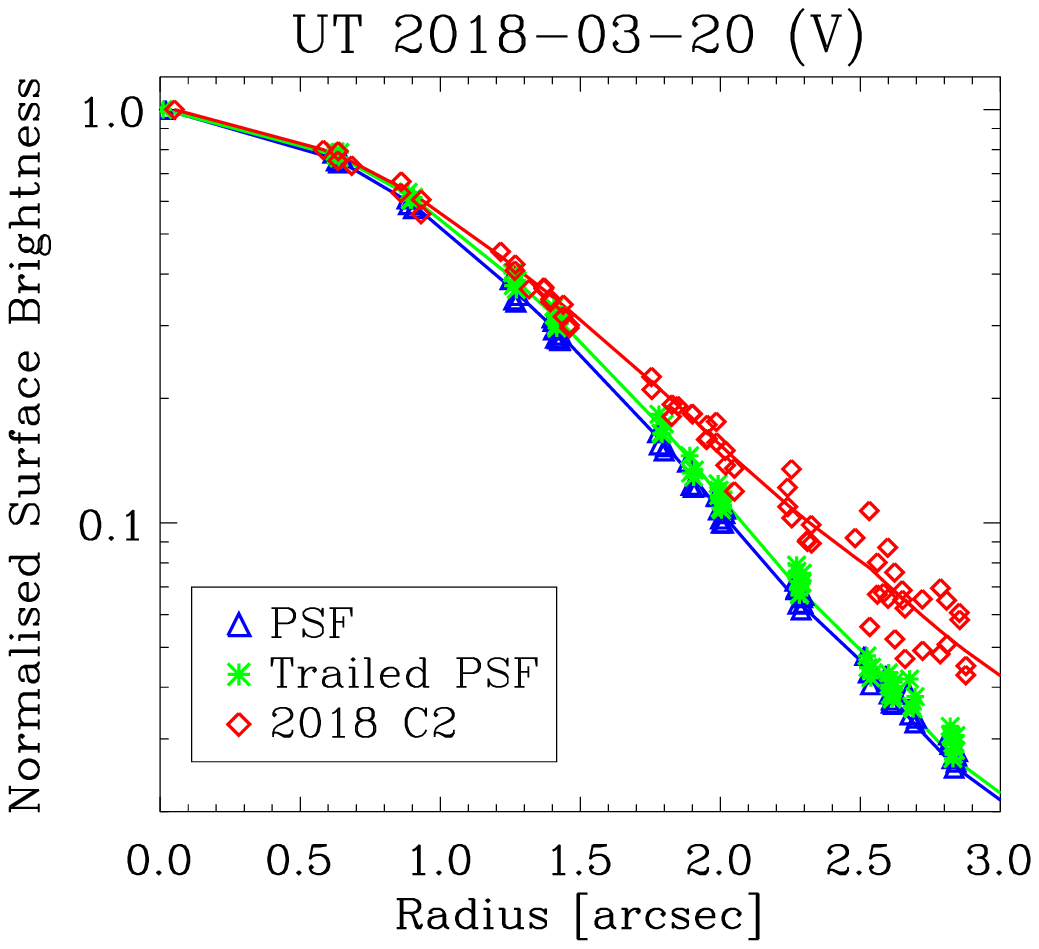} \\
    \centering\small ~~~(c)
  \end{tabular}
  \caption{
The logarithmic surface brightness of the PSF, the trailed PSF, and 2018 C2 from UT 2019 March 18--20, when the images had the best quality because of the weather, is plotted versus apparent radial distance from their peak pixels. Each profile is normalized to unity at the the peak. Solid lines in the three panels are azimuthal mean values from the corresponding pixel counts, in the same colors as for the symbols. Despite the scatter, a deviation from the radial profile of the trailed PSF can be readily seen for that of 2018 C2, thus suggesting a low activity level.
  \label{fig_18C2_radp}
  }
\end{figure}

Our examination of individual images in which seeing was good and 18C2 was not close to any of the background sources revealed that the azimuthally-averaged FWHM of 18C2 is consistently larger than the values of the field stars by $\gtrsim$0\arcsec.1. This suggests a low level of activity of the object. To further verify this, we compared the star profiles in the NEXT images and the profile of 18C2 in the coadded images. The {\it V}-band images were examined, for the sake of the bandpass covering both typical gas and dust components of a potential coma at $r_{\rm H} \approx 2$ AU. We then extracted the PSF profiles from the field stars, with its peak normalized to unity. To make sure the tracking of the telescope, which, as mentioned in Section \ref{sec_obs}, was in the sidereal rate, did not contribute the observed widening of the PSF profile, we convolved the PSF with a motion vector, whose length is determined by the apparent distance covered by 18C2 during the exposure time. The corrected profile of the PSF indeed has a shallower slope, but far from enough to account for the wider profile of 18C2 (Figure \ref{fig_18C2_radp}). Thus, although only a possible extremely faint coma was observed in the images, it is more than confident that 18C2 has a very low level of activity.

\begin{figure}
  \centering
  \begin{tabular}[b]{@{}p{0.3\textwidth}@{}}
    \centering\includegraphics[width=0.36\textwidth]{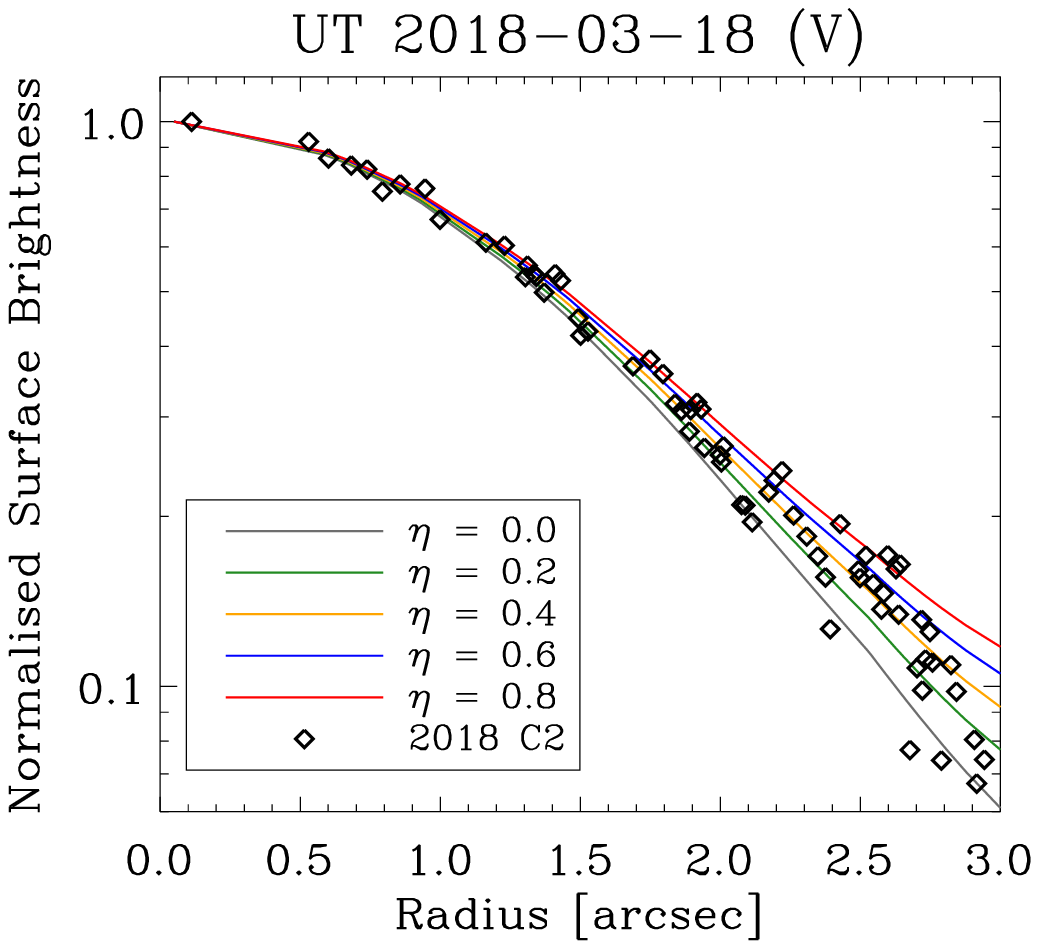} \\
    \centering\small ~~~(a)
  \end{tabular}%
  \quad
  \begin{tabular}[b]{@{}p{0.3\textwidth}@{}}
    \centering\includegraphics[width=0.36\textwidth]{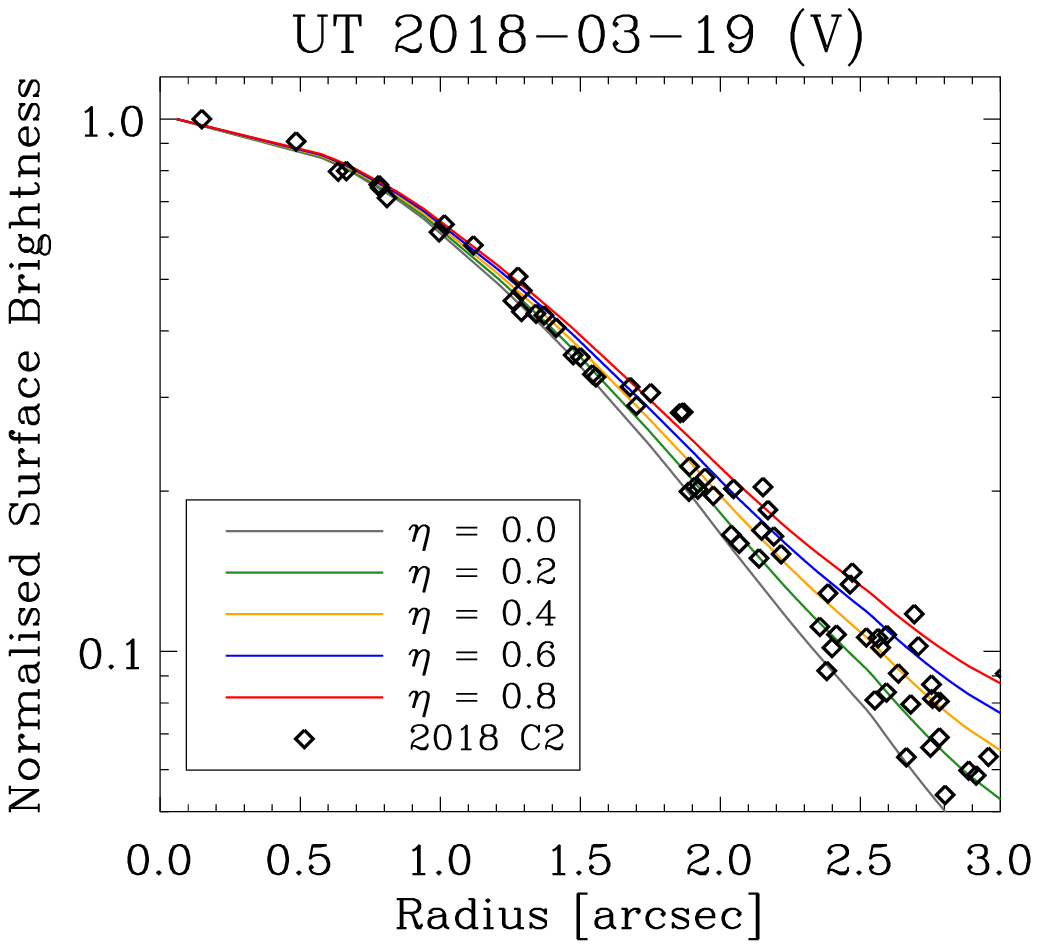} \\
    \centering\small ~~~(b)
  \end{tabular}
  \quad
  \begin{tabular}[b]{@{}p{0.34\textwidth}@{}}
    \centering\includegraphics[width=0.36\textwidth]{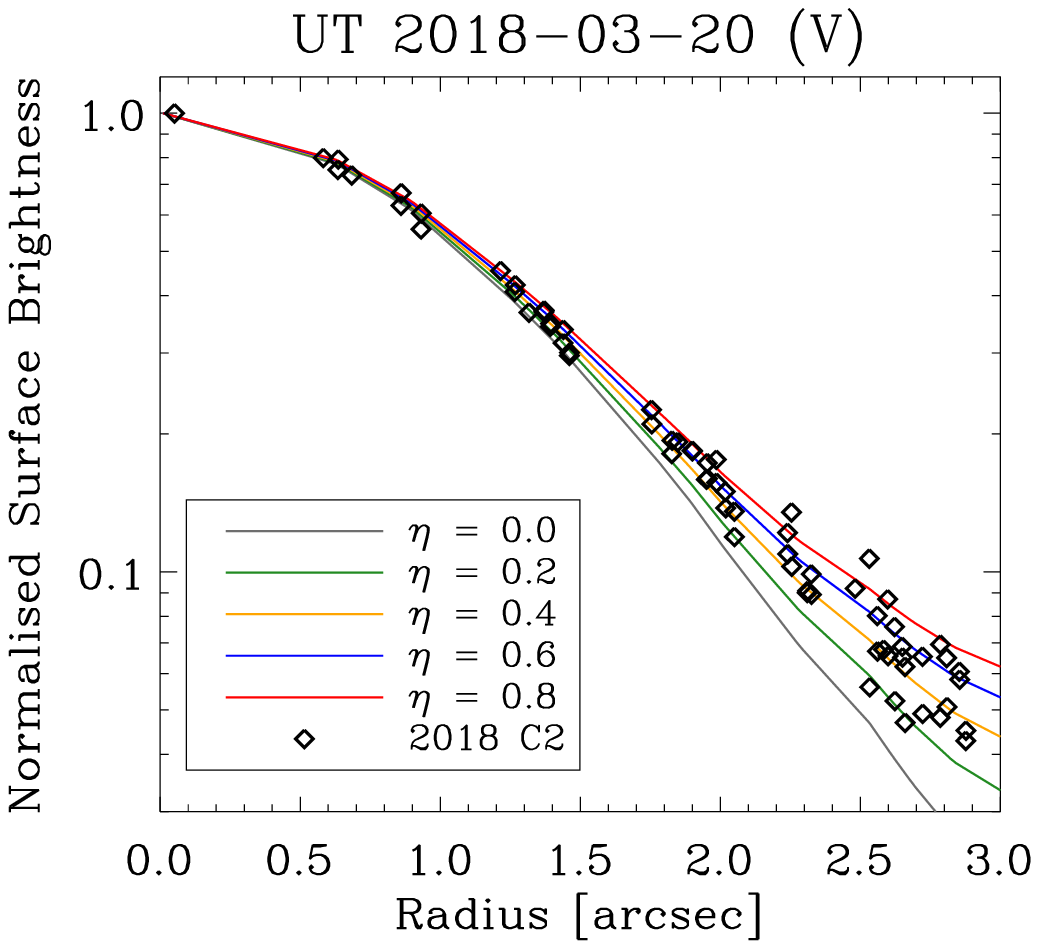} \\
    \centering\small ~~~(c)
  \end{tabular}
  \caption{
The normalized surface brightness profile of 2018 C2 from UT 2019 March 18--20, compared with models of different levels of coma contribution in a steady state, in terms of $\eta$, the ratio between the signal of the coma and that of the nucleus within angular radius $\varrho_0 = 5\arcsec.7$. Profiles of a bare nucleus correspond to the models with $\eta = 0$, which are essentially the profiles of the trailed PSF in Figure \ref{fig_18C2_radp}. By visual comparison, models with $\eta \approx 0.4$ generally match the observed radial profile of the target the best for all the three nights.
  \label{fig_18C2_radp_mod}
  }
\end{figure}

In order to quantitatively obtain the activity level of 18C2, we employed a method similar to the one by Luu \& Jewitt (1992). First, a series of steady coma models of different activity levels, parameterized by $\eta$, which is the ratio between the coma flux and the nucleus flux enclosed by a common aperture radius $\varrho_0$ at the time of observations, were generated. We then convolved the models with the trailed PSF. Finally, a nucleus represented by the trailed PSF was added to the models at the coma center. The profile of such a synthesized comet, denoted as $I$, under the polar coordinates $(\varrho, \theta)$ is thereby
\begin{equation}
I\left( \varrho, \theta \right) = k_{\rm N}  \left[ \delta\left(\varrho \right) + \frac{\eta}{\varrho_0 \varrho} \int\limits_{0}^{\varrho_{0}} \varrho \delta \left(\varrho \right) {\rm d} \varrho \right] \ast \mathcal{P} 
\label{eq_profile},
\end{equation}
\noindent in which $k_{\rm N}$ is an arbitrary scaling factor of the nucleus, $\delta$ is the Dirac delta function, $\mathcal{P}$ is the trailed PSF with the total flux normalized to unity, and the symbol $\ast$ is the convolution operation. We chose $\varrho_0 = 5\arcsec.7$ to be in line with the photometric aperture size. The convolution was performed at a resolution eleven times finer than the original, so as to avoid an effect due to the potentially inaccurate replacement of the singularity at $\varrho = 0$ by numerical integration, and then rebinned back to the original resolution. Results are shown in Figure \ref{fig_18C2_radp_mod}. We can see that the models with $\eta \approx 0.4 \pm 0.1$ appear to match the observed profile of 18C2 the best by visual inspection. After converting the observing geometry to $\alpha = 0\degr$, we then obtain $\eta_0 = \eta \Phi_{\rm N} \left( \alpha \right) / \Phi_{\rm C} \left(\alpha \right) \approx 0.3 \pm 0.1$.

The mean mass-loss rate of 18C2 can be estimated in a manner that the coma is produced in the diaphragm crossing time. Therefore, it can be written as
\begin{equation}
\overline{\dot{M}}_{\rm N} = \frac{4 \pi \eta_{0} \rho \bar{\mathfrak{a}} R_{\rm N}^{2} v}{3 \varrho_0 {\it \Delta}}
\label{eq_mloss}.
\end{equation}
\noindent Here, $\rho$ is the bulk density of the dust grains, $\bar{\mathfrak{a}}$ is the mean grain radius, $v$ is the speed of the dust grains. Unfortunately, low SNR of the observed coma of the object prohibits dust analysis, so none of the three aforementioned parameters are known. We have to assume $\rho = 10^{3}$ kg m$^{-3}$, $\bar{\mathfrak{a}} = 0.5$ \micron, which is the peak wavelength of the {\it V}-band filter, and that $v$ varies proportional to $r_{\rm H}^{-1/2}$ (Delsemme 1982). Substituting these numbers, Equation (\ref{eq_mloss}) yields $\overline{\dot{M}}_{\rm N} \approx 0.7 \pm 0.2$ kg s$^{-1}$, with all the uncertainties taken into consideration. This is no match to mass-loss rates of the long-period comets (e.g., $\ga$10$^{5}$ kg s$^{-1}$ for C/1995 O1 (Hale-Bopp), and $\sim$10$^{2}$ kg s$^{-1}$ for C/2011 L4 (PANSTARRS); Jewitt \& Matthews 1999; Yang et al. 2014), but similar to the activity levels of active asteroids (e.g., 0.7--1.6 kg s$^{-1}$ for 133P/(7968) Elst-Pizarro, and 0.2--0.4 kg s$^{-1}$ for 313P/Gibbs; Jewitt et al. 2015 and citations therein). In fact, this classifies 18C2 as one of the least active objects hitherto known from the Oort cloud, together with 17U7, given that no activity has been seen so far.

Provided that the activity is due to sublimation of some volatile, the fraction of active area,
\begin{align}
\nonumber
\zeta & \equiv \frac{\overline{\dot{M}}_{\rm N}}{4 \pi R_{\rm N}^{2} f_{\rm s}} \\
 & = \frac{\eta_0 \rho \bar{\mathfrak{a}} v}{3 f_{\rm s} \varrho_{0} {\it \Delta}}
\label{eq_faa},
\end{align}
\noindent where $f_{\rm s}$ is the mass flux of the sublimating substance, can be estimated. In Equation (\ref{eq_faa}), the dust-to-gas production rate is assumed to be unity. One needs to solve the energy equilibrium equation for $f_{\rm s}$:
\begin{equation}
\left( 1 - A_{\mathrm{B}} \right) \frac{S_{\odot}}{r_{\mathrm{H}}^2} = \chi \left[ \epsilon \sigma T^4 + L\left( T \right) f_{\mathrm{s}} \left( T \right) \right]
\label{eq_enrgy},
\end{equation}
\noindent where $A_{\mathrm{B}}$ is the Bond albedo, $S_{\odot} = 1361$ W m$^{-2}$ is the solar constant at $r_{\mathrm{H}} = 1$ AU, $1 \le \chi \le 4$ is a dimensionless coefficient factor with the two ends corresponding to subsolar and isothermal scenarios, respectively, $\epsilon$ is the emissivity, $\sigma = 5.67 \times 10^{-8}$ W m$^{-2}$ K$^{-4}$ is the Stefan-Boltzmann constant, $T$ is the surface temperature in K, $L$ (in J kg$^{-1}$) is the latent heat of the sublimating ice, which, given the small heliocentric distance, is likely water ice in this case. Heliocentric distance is expressed in AU. Taking $A_{\rm B} = 0.01$ and $\epsilon = 0.9$, at $r_{\rm H} \approx 2.2$ AU, Equation (\ref{eq_enrgy}) yields $6 \times 10^{-6} \la f_{\rm s} \la 8 \times 10^{-5}$ kg s$^{-1}$ m$^{-2}$, the two limits corresponding to the isothermal and subsolar cases. Plugging the numbers in, we obtain $\zeta \sim 10^{-5}$--10$^{-4}$ as the fraction of active area of 18C2. Once again, it confirms that its activity is significantly weaker than that of typical long-period comets, and even weaker than many of the Jupiter-family comets by at least two orders of magnitude, such as 67P/Churyumov-Gerasimenko, which has a few percent (Combi et al. 2012), but stronger than Ceres ($\zeta < 10^{-5}$; K{\"u}ppers et al. 2014).

We have no access to physical observations of 17U7 so we cannot quantitatively determine its activity level. However, there is evidence that it is also weakly active, in that it exhibited a slight asymmetry suggesting a very short eastward tail in Pan-STARRS images from 2017 November, in which its FWHM is larger than nearby stars' by $\sim$12\% (R. Weryk, private communication).

\section{\uppercase{Orbits}}
\label{sec_orb}

Astrometric measurements of 17U7 and 18C2 were retrieved from the MPC. The former are comprised of 87 data points spanning from UT 2017 August 18 to 2018 February 10, and the latter contains 460 data points covering an observing arc from UT 2018 January 28 to April 26. Additional high-quality astrometry of 18C2 was measured in the NEXT images, and from the Lulin Observatory.

The astrometric measurements of both objects were then debiased based upon Farnocchia et al. (2015), and weighted using a scheme devised by Vere{\v s} et al. (2017). We then fed the data to an orbit determination code, \textit{EXORB9}, written by A. Vitagliano as a part of the \textit{SOLEX12} package, to obtain optimized solutions to the orbits of 17U7 and 18C2, with inclusion of gravitational perturbations from the planets, Pluto, and the most massive 27 asteroids in the main belt, and relativistic corrections for completeness. During the process, we searched for ad hoc outliers which were no less than three times the assigned positional uncertainties, but found none for both objects. We then obtained the J2000 heliocentric equatorial Cartesian state vectors at epochs of their respective first available observations, which were then converted to the heliocentric ecliptic orbital elements. The results are presented in Tables \ref{tab_orb_17U7} and \ref{tab_orb_18C2}, where comparisons to the values given by JPL HORIZONS are shown. The differences in the orbital elements result from different weighting schemes and force models (and also different datasets for 18C2), but are within the corresponding uncertainty levels. 

Nevertheless, the result clearly shows that, at the moment, both 17U7 and 18C2 have barely heliocentric hyperbolic orbits around the Sun with $e = 1.002$.

\section{\uppercase{Dynamical Evolution}}
\label{sec_dyn}

Although 17U7 and 18C2 have heliocentric hyperbolic orbits, the eccentricities are both much smaller than that of `Oumuamua ($e \approx 1.2$), and by no means are they outstanding compared to some hyperbolic comets, e.g., $e \approx 1.06$ for C/1980 E1 (Bowell), which is a solar system object but was ejected due to a close approach to Jupiter (Buffoni et al. 1982). For the majority of the comets in hyperbolic trajectories when close to the Sun, their original orbits are found to be elliptical before encountering accelerating planetary perturbations (Kres\'ak 1992). Other comets with even higher eccentricities can be explained by the nongravitational effects caused by anisotropic mass loss (Kr{\'o}likowska 2001). These comets tend to be more active than short-period comets, and commonly exhibit activity even at heliocentric distances $r_{\mathrm{H}} \gtrsim 10$ AU from the Sun preperihelion, e.g., C/2010 U1 (Boattini), C/2017 K2 (PANSTARRS; Hui et al. 2018). 17U7 and 18C2, however, appear to be much less active. Heliocentric hyperbolic objects of similar refractory levels have never been observed before. Therefore, in order to have a better comprehension about their dynamical histories, we performed analysis of their orbital dynamics.

We generated 1,000 clones for both 17U7 and 18C2 so as to encompass uncertainties in our orbit determination. Initial conditions were generated from the covariance matrices with offsets added to the nominal values obtained in Section \ref{sec_orb}, which were scaled by random numbers following Gaussian distributions having variances equal to the corresponding eigenvalues. We then employed our modified version of the \textit{MERCURY6} package\footnote{The original version of the package can be downloaded from \url{http://star.arm.ac.uk/~jec/mercury/mercury6.tar}.} (Chambers 1999) and conducted N-body integration of the clones. The hybrid symplectic/Bulirsch-Stoer integrator algorithm, which is capable of handling close encounters (Eggl \& Dvorak 2010), if any, was utilised for integration over time intervals covering previous and next perihelion passages of each clone, provided that the orbits are not hyperbolic in the reference system of the solar system barycentre. An initial step size of 5 days was adopted. Also included are gravitational perturbations from all the eight major planets, relativistic corrections, and the galactic tide. Influences from the galactic tide are negligible when objects are in the planetary region where perturbations from the gas giants are dominant, but end up being a major perturber when they are $\gtrsim 10^3$ AU from the Sun, i.e., in the Oort cloud (see Dones et al. 2004 and references therein). For sake of computation speed, gravitational perturbations from Pluto and the most massive asteroids in the main belt were discarded, as during our test in Section \ref{sec_orb}, their effects were found to be almost undetectable.

Note that potential stellar perturbations were not considered in the integration. The reason is that majority of the passing stars of potential encounters are low-mass dwarfs (Garc{\'{\i}}a-S{\'a}nchez et al. 2001), and consequently, the perturbation from the galactic tide is orders-of-magnitude stronger than perturbations from passing stars. The statistics are therefore not expected to change much due to these induced random kicks, despite that the uncertainty in this work may be slightly underestimated. Indeed, Dybczy{\'n}ski (2006) has shown that no known stars had significant influences on motions of long-period comets recently within the last $\sim$10 Myr, nor in a future similar time frame, justifying the omission. The controversial distant giant planet (Trujillo \& Sheppard 2014; Batygin \& Brown 2016) was not considered in the integration either.

\begin{figure}
\epsscale{1.0}
\begin{center}
\plotone{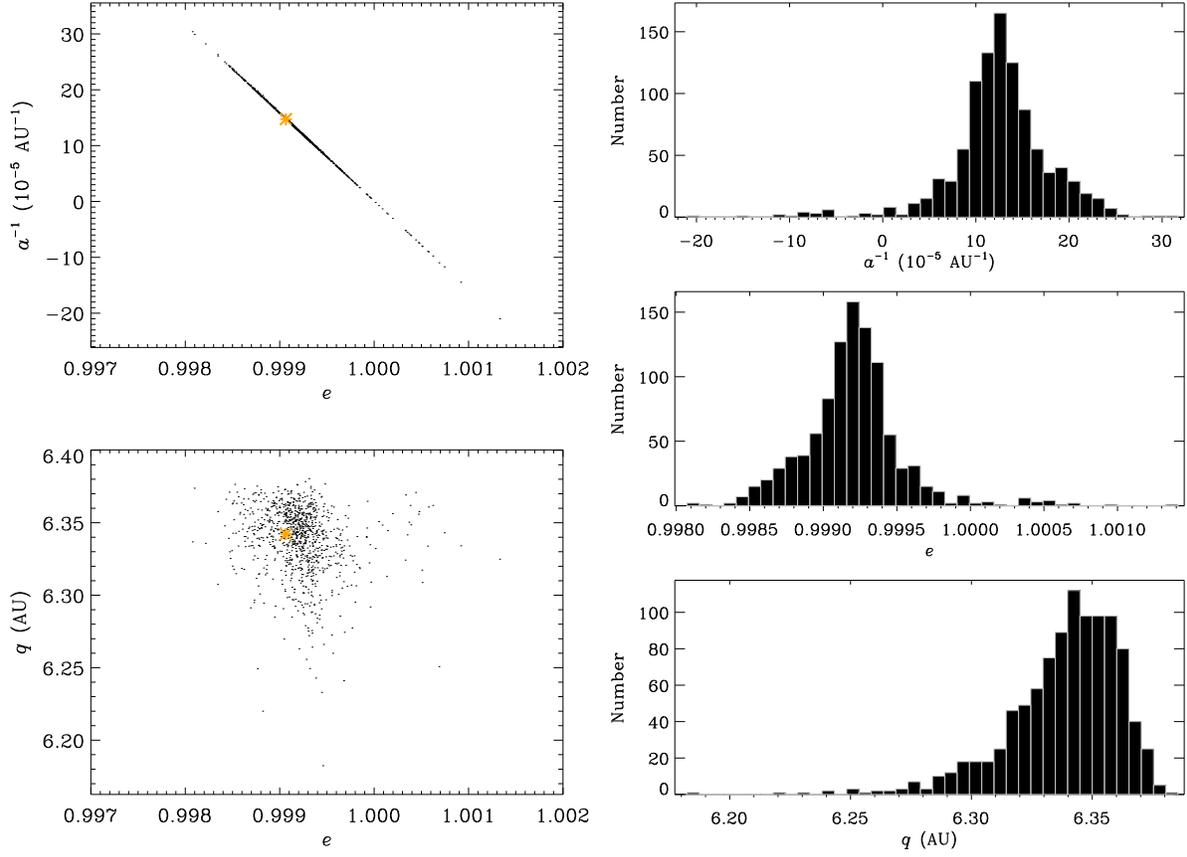}
\caption{
Statistics of the solar system barycentric orbital elements of the 1,000 Monte Carlo clones together with the nominal orbit of 2017 U7 at the previous perihelion passage, projected onto the $a^{-1}$-$e$ and $q$-$e$ space. Its previous perihelion occurred at an epoch $t_{\rm p} \in (-1.3, -0.5)$ Myr from J2000, with median and standard deviation as $\tilde{t}_{\rm p} = -0.75 \pm 0.10$ Myr. The nominal orbit is plotted as an orange asterism in the left two panels. Note that none of the distributions of the selected orbital elements are Gaussian.
\label{fig_17U7_distpp}
} 
\end{center} 
\end{figure}

\begin{figure}
\epsscale{1.0}
\begin{center}
\plotone{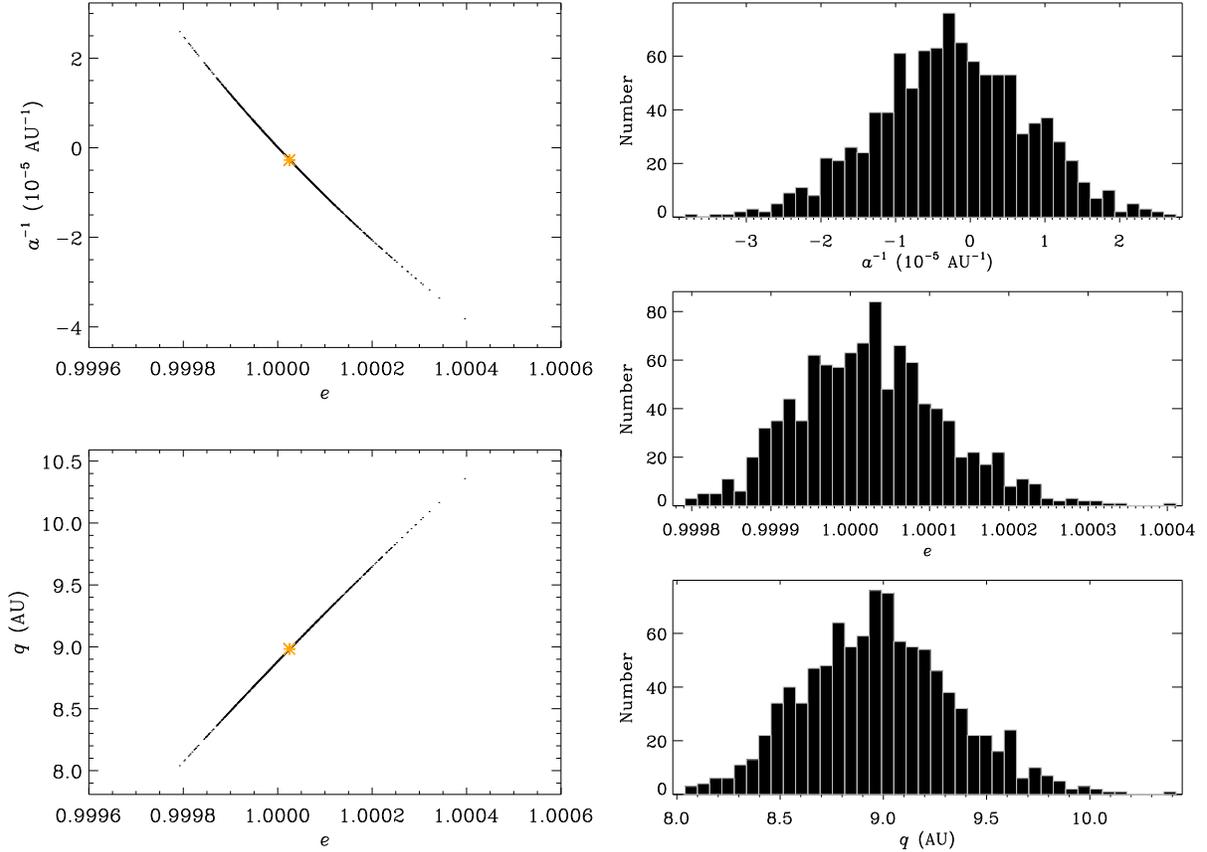}
\caption{
Statistics of the solar system barycentric orbital elements of the 4,000 Monte Carlo clones of 2017 U7 at epoch 1.5 Myr from J2000, projected onto the $a^{-1}$-$e$ and $q$-$e$ space. The nominal orbit is plotted as an orange asterism in the left two panels. The median values of the reciprocal semimajor axis and the perihelion distance are $\tilde{a}^{-1} = \left(-2.6 \pm 10.2 \right) \times 10^{-6}$ AU$^{-1}$, and $\tilde{q} = 8.98 \pm 0.36$ AU, at epoch of 1.5 Myr from J2000, when the object is $r_{\rm H} \ga 10^{4}$ AU from the Sun.
\label{fig_17U7_distnp}
} 
\end{center} 
\end{figure}

We backward and forward integrated the clones of 17U7. Our simulation result shows that it is most likely a dynamically old object from the Oort cloud, as all of the clones were found to have formerly entered the region of $r_{\rm H} < 15$ AU, a threshold value for significant planetary perturbations,\footnote{Thus the division line between dynamically old and new comets is commonly set at previous perihelion distance $q_{|} = 15$ AU (e.g., Dybczy{\'n}ski 2001; Kr{\'o}likowska \& Dybczy{\'n}ski 2010).} during the previous perihelion, which occurred at an epoch $-1.3 \la t_{\rm p} \la -0.5$ Myr from J2000, with its median and standard deviation as $\tilde{t}_{\rm p} = -0.75 \pm 0.10$ Myr. Statistics of the orbits of the clones at the previous perihelion are shown in Figure \ref{fig_17U7_distpp}. Only $\sim$2\% of the clones in heliocentric hyperbolic orbits at the previous perihelion means that it is highly unlikely that 17U7 is a recent interstellar interloper like `Oumuamua (de la Fuente Marcos \& de la Fuente Marcos 2017). However, it has a likelihood of becoming an interstellar object after the upcoming perihelion passage, because in our simulation for the next 1.5 Myr, $\sim$60\% of the clones, including the nominal orbit, become unbound to the solar system (see Figure \ref{fig_17U7_distnp}), although none of them have strongly hyperbolic eccentricities ($e \lesssim 1.0004$) like that of `Oumuamua. Its eccentricity will be boosted immediately after the forthcoming perihelion in 2019 September and a further kick due to a moderate close approach to Jupiter in 2020 May ($\sim$1.66 AU), and the perihelion distance will monotonously increase to $q \gtrsim 8$ AU on the outbound leg, by the epoch of 1.5 Myr from J2000.

\begin{figure}
\epsscale{1.0}
\begin{center}
\plotone{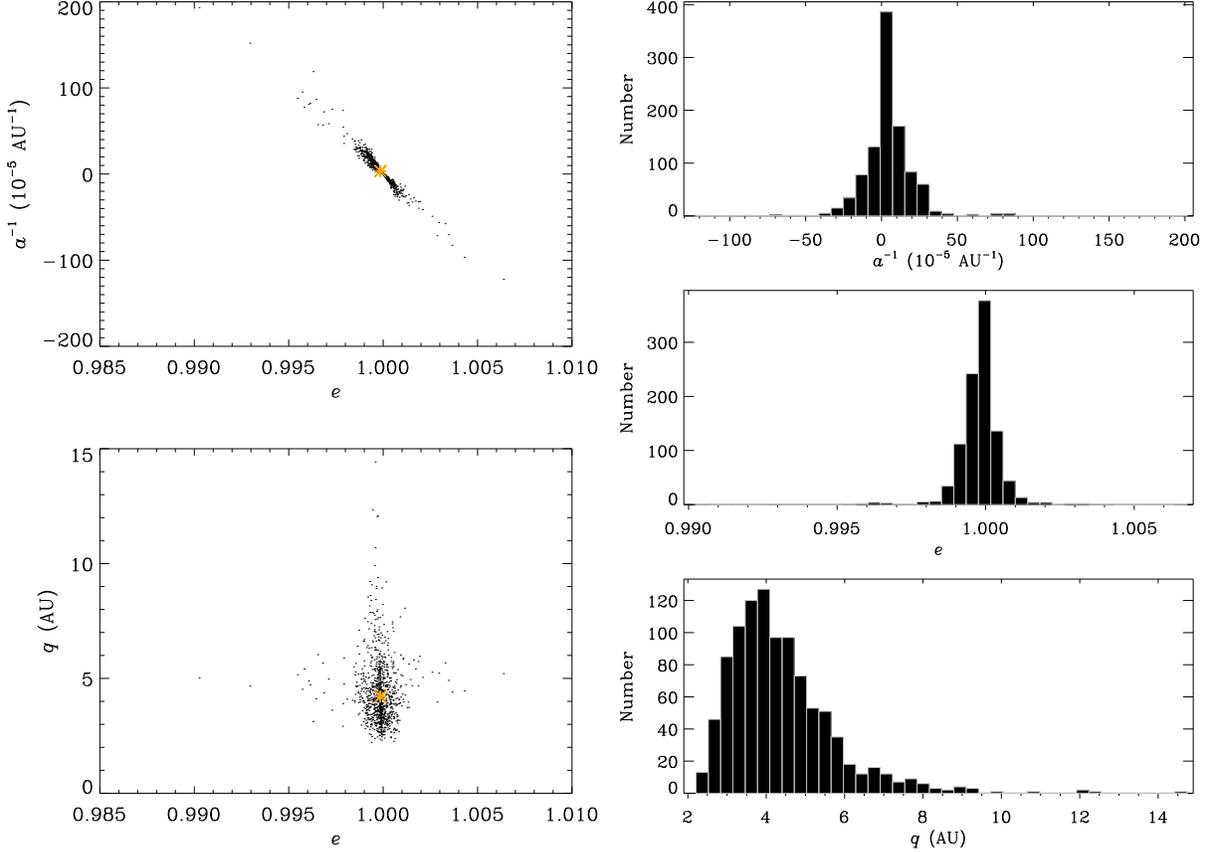}
\caption{
Distributions of the barycentric orbital elements of the Monte Carlo clones of 2018 C2 in the $a^{-1}$-$e$ and $q$-$e$ space at the previous perihelion ($t_{\rm p} \in (-6.3, -3.9)$ Myr from J2000, with median and standard deviation as $\tilde{t}_{\rm p} = -4.72 \pm 0.34$ Myr). Two of the clones which have $e = 0.874$ and 1.015, respectively, and are distinctly separated from the majority are not shown in the left two panels and the histograms for $a^{-1}$ and $e$ for visuality. The orange asterism marks the nominal orbit in the left two panels. At the previous perihelion passage, there is a 29\% chance that 2018 C2 had a barycentric eccentricity $e > 1$. All of the clones are found to have previously entered the solar system with $r_{\rm H} < 15$ AU. 
\label{fig_18C2_distpp}
} 
\end{center} 
\end{figure}

\begin{figure}
\epsscale{1.0}
\begin{center}
\plotone{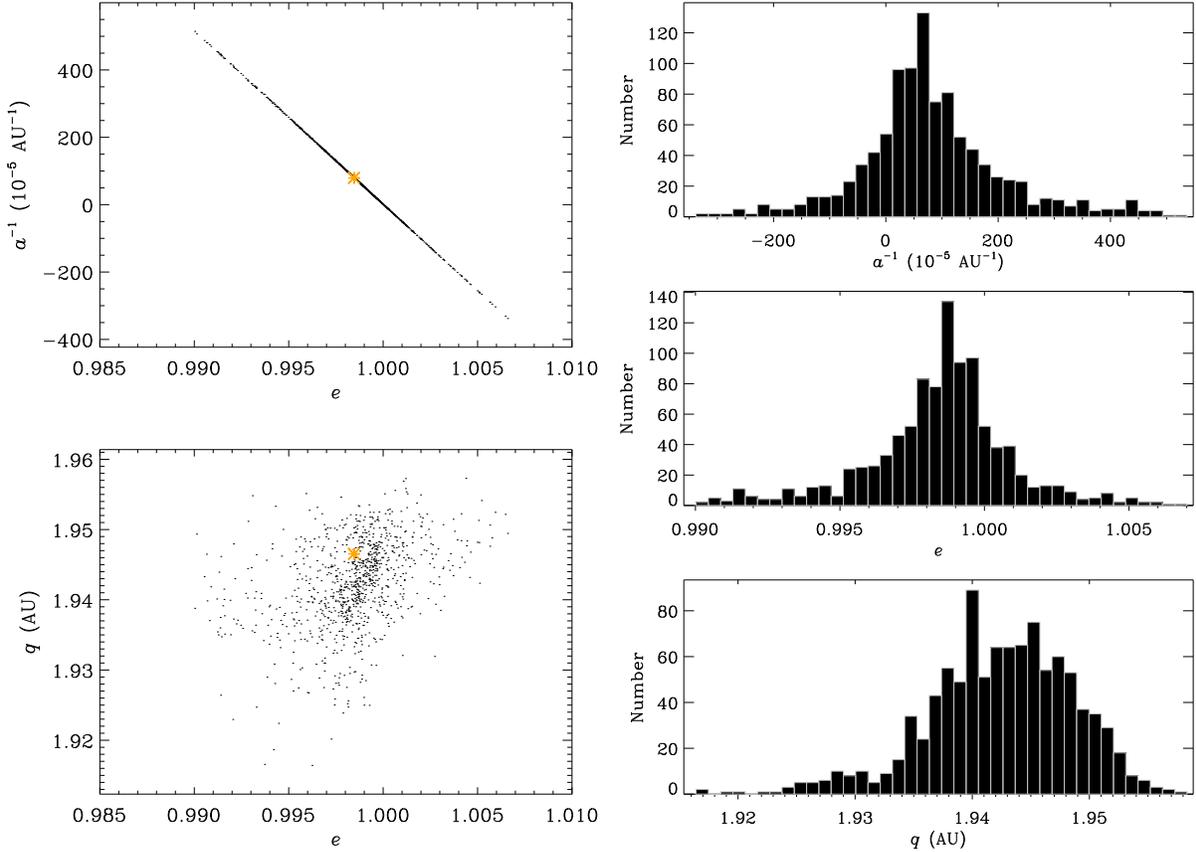}
\caption{
The selected barycentric orbital elements of the 1,000 Monte Carlo clones of 2018 C2 at the next perihelion passage, which will occur at an epoch $t_{\rm p} \in (0.13, 0.14)$ Myr from J2000 (with median and standard deviation as $\tilde{t}_{\rm p} = 133.3 \pm 0.9$ kyr), in the $a^{-1}$-$e$ and $q$-$e$ space. The nominal orbit is marked by an orange asterism in the left two panels. It is highly confident that if 2018 C2 is able to survive at all, it will enter the inner solar system again after the forthcoming perihelion, with a perihelion distance similar to the current one, $q \in [1.92, 1.96]$ AU. A fraction of 20\% of the clones have barycentric eccentricity $e > 1$, which means that there is a moderate chance for 2018 C2 to escape the solar system in the future.
\label{fig_18C2_distnp}
} 
\end{center} 
\end{figure}

We also performed backward and forward integration of the clones of 18C2. Statistics of the selected barycentric orbital elements at previous perihelion, which took place at an epoch of $-6.3 \la t_{\rm p} \la -3.9$ Myr from J2000, are plotted in Figure \ref{fig_18C2_distpp}. We can see that it is almost certain that 18C2 is a dynamically old object from the Oort cloud as well, since all of the clones have previous perihelion distance $q < 15$ AU. To be more exact, 996 of the clones have previously passed perihelion closer than 10 AU, and in particular, 794 of them have even been inside the orbit of Jupiter. After the current perihelion, its barycentric semimajor axis will significantly shrink as a result of planetary perturbations; all the clones have their subsequent perihelion passages at epochs of $0.13 \la t_{\rm p} \la 0.14$ Myr from J2000, with $1.92 \le q \le 1.96$ AU, similar to the current perihelion distance. Therefore, provided that 18C2 can survive at all, it will then pay yet another visit to the inner solar system. Distributions of the heliocentric orbital elements for the next perihelion are shown in Figure \ref{fig_18C2_distnp}. We notice that 199 of the clones (20\% of the total) have their barycentric eccentricity barely greater than 1 ($e < 1.007$), meaning that there is a chance that 18C2 is potentially able to escape the solar system in the further future.


\section{\uppercase{Discussion}}
\label{sec_disc}

Why are 17U7 and 18C2 distinctly less active than typical long-period comets? Dynamically, as Section \ref{sec_dyn} shows, the two objects do not exhibit any atypical properties in comparison to the comets from the same source within the examined time interval. It has been well known that the observed number of the Oort-spike comets is far more than expected based on the number of dynamically old comets, which is thought to be due to physical evolution, including disintegration and exhaustion of volatiles (Oort 1950). Thus, one of the possibilities is that 17U7 and 18C2 are thermally evolved (dormant and extinct included) long-period comets. Indeed, our dynamical simulations suggest that both objects are unlikely dynamically new so they have previously visited the planetary region. Particularly, for 18C2, its color consistent with previously known long-period comets appears to strengthen the idea that it is likely a thermally evolved comet. 

Another possibility is that 17U7 and 18C2 might have been inner solar system objects in the past that is beyond the time coverage of our dynamical integration, such that most of the volatiles but subterranean ones have been depleted, and were subsequently scattered to the Oort cloud region by the giant planets, similar to what is suggested for comet C/2014 S3 (PANSTARRS; Meech et al. 2016). The observed flux of long-period comets with perihelion distances no less than that of 18C2 is $\sim$20 yr$^{-1}$ (from the statistics of discoveries of long-period comets in the past ten years). The fraction of refractory objects in the Oort cloud is estimated to be $\sim$1\% (Weissman \& Levison 1997; Shannon et al. 2014), this is translated to an observed flux of such objects as $\sim$0.1 yr$^{-1}$. Given the observed statistics based on merely a few samples, the flux is probably too small, thereby making the argument that the two objects are originally from the inner solar system less favourable. However, with a caveat that the dynamical models might not have disclosed the actual fraction of Oort-cloud asteroids, and also nowadays the all-sky surveys have ever-better capability of detecting faint objects, this possibility cannot be fully refuted.


\section{\uppercase{Summary}}
\label{sec_sum}

Key conclusions of our study about the two hyperbolic objects 2017 U7 and 2018 C2 are summarized as follows.

\begin{enumerate}

\item Both are most likely dynamically old objects that have previously entered the planetary region of the solar system. In the subsequent 1.5 Myr, 2017 U7 has a $\sim$60\% likelihood that it will be unbound to the solar system after the observed forthcoming perihelion as a result of a moderate close approach to Jupiter in 2020 May. 2018 C2 will revisit the inner solar system 0.13--0.14 Myr later, with perihelion distance $q \approx 2$ AU, if it is able to physically survive at all.

\item Based on observations from Xingming Observatory, the color of 2018 C2 resembles those of the long-period comets, Trojans, and D-type asteroids: $m_B - m_V = 0.75 \pm 0.03$, $m_V - m_R = 0.41 \pm 0.02$, and $m_R - m_I = 0.37 \pm 0.03$. Assuming geometric albedo $p_R = 0.04$, its effective radius is $4.4 \pm 0.5$ km. 

\item A possible extremely faint coma of $\sim$9\arcsec~across was observed for 2018 C2 in the coadded images from Xingming Observatory. Its radial profile confirms that it has a very low level of activity, significantly weaker than typical long-period comets, and even many of the Jupiter-family comets. The mass-loss rate is estimated to be $\overline{\dot{M}}_{\rm N} \approx 0.7 \pm 0.2$ kg s$^{-1}$, corresponding to the fraction of active area $\sim$10$^{-5}$--10$^{-4}$. 

\item The dynamical simulation and the observed color of 2018 C2 support that it is probably a thermally evolved comet. Similarly, 2017 U7 may be a comet of this type as well, given its dynamical history and appearance in the Pan-STARRS data. Yet the possibility of their origin in the inner solar system, although less favourable, cannot be fully rejected.

\end{enumerate}

\acknowledgements
{
We heartedly thank Xing Gao, the director and founder of Xingming Observatory, for conducting observations of 2018 C2 from the observatory at our request. We shall also thank a hacker who did not zap Xingming's telescope control computer earlier but two days later after the observations of 2018 C2 were finished. The operation of Xingming Observatory is made possible by the generous support from the Xinjiang Astronomical Observatory of the Chinese Academy of Sciences. The Ningbo Bureau of Education and Xinjiang Observatory Telescope is sponsored by Ningbo Bureau of Education. We appreciate Aldo Vitagliano for debugging and modifying the {\it SOLEX12} package under our request, Robert Weryk for inspecting the Pan-STARRS images of 2017 U7, and Lulin Observatory, National Central University, for providing us with good-quality astrometry of 2018 C2, aided by Chi-Hao Hsia. Discussions with Hilke Schlichting and Quan-Zhi Ye have benefited this work. Comments from Ariel Graykowski, David Jewitt and the anonymous referee helped improve the manuscript. Observers who submitted good quality astrometry are acknowledged. This work is funded by a grant from NASA to David Jewitt.
}

\clearpage

\appendix
\section{\uppercase{Derivation of Compound Phase Function for Comets}}
\label{apndx}
Flux from a comet is contributed by both its nucleus and coma. Here we only consider its scattering flux of the sunlight. The total flux of the comet at phase angle $\alpha$ is
\begin{align}
\nonumber
F \left( \alpha \right) & = F_{\rm N} \left(\alpha \right) + F_{\rm C} \left(\alpha \right) \\
 & = F_{\rm N} \left(0\right) \Phi_{\rm N} \left(\alpha \right) + F_{\rm C} \left(0\right) \Phi_{\rm C} \left(\alpha \right)
\label{eq_A1}.
\end{align}
\noindent Using the relationship $\eta_0 = F_{\rm C} \left(0\right) / F_{\rm N}\left(0\right)$, Equation (\ref{eq_A1}) then becomes
\begin{equation}
F \left( \alpha \right) = F_{\rm N} \left(0 \right) \left[\Phi_{\rm N} \left(\alpha\right) + \eta_0 \Phi_{\rm C} \left(\alpha\right) \right]
\label{eq_A2}.
\end{equation}
\noindent Let us write the left-hand side of Equation (\ref{eq_A2}) in the same form as $F\left(\alpha\right) = F\left(0\right) \Phi \left(\alpha\right)$, whereby we obtain
\begin{equation}
F\left(\alpha\right) = F_{\rm N} \left(0 \right) \Phi \left(\alpha \right) \left(1 + \eta_0 \right)
\label{eq_A3},
\end{equation}
\noindent after $\eta_0$ is introduced. Combining Equations (\ref{eq_A2}) with (\ref{eq_A3}), one can solve for $\Phi\left(\alpha\right)$ as
\begin{equation}
\Phi \left(\alpha \right) = \frac{\Phi_{\rm N} \left(\alpha \right) + \eta_0 \Phi_{\rm C} \left(\alpha \right)}{1 + \eta_0}
\label{eq_A4}.
\end{equation}
\noindent The result makes sense as the compound phase function is in essence the weighted mean on the phase functions of the nucleus and coma. If the flux of a comet is dominated by its dust coma, i.e., $\eta \gg 1$, we obtain $\Phi \approx \Phi_{\rm C}$. Otherwise if the coma is almost absent, i.e., $\eta \to 0$, Equation (\ref{eq_A4}) yields $\Phi \approx \Phi_{\rm N}$, which means that the phase function is mainly determined by the one of the nucleus.

\clearpage{}

\begin{deluxetable}{cccccccc}
\tablecaption{Observing Geometry of 2018 C2 from Xingming Observatory
\label{tab_geo}}
\tablewidth{0pt}
\tablehead{ 
\colhead{Date \& Time (UT)} & \colhead{$t_{\rm exp}$ (s)\tablenotemark{[1]}} & \colhead{$r_{\rm H}$ (AU)\tablenotemark{[2]}}  & 
\colhead{$\it \Delta$ (AU)\tablenotemark{[3]}} & \colhead{$\alpha$ (\degr)\tablenotemark{[4]}} & 
\colhead{$\varepsilon$ (\degr)\tablenotemark{[5]}} & \colhead{$X$\tablenotemark{[6]}} &
\colhead{Condition\tablenotemark{[7]}}
}
\startdata

2018 Mar 11 18:03--22:12 & 60 & 2.199 & 1.538 & 23.3 & 119.1 & 1.37--1.00 & C \\
2018 Mar 13 21:12--22:38 & 60 & 2.187 & 1.520 & 23.3 & 119.5 & 1.01--1.01 & C \\
2018 Mar 18 20:23--22:40 & 90 & 2.162 & 1.480 & 23.4 & 120.4 & 1.03--1.02 & -- \\
2018 Mar 19 19:20--22:38 & 90 & 2.157 & 1.473 & 23.4 & 120.6 & 1.10--1.03 & -- \\
2018 Mar 20 20:08--22:39 & 90 & 2.152 & 1.465 & 23.4 & 120.7 & 1.04--1.03 & -- \\

\enddata

\tablenotetext{[1]}{Individual exposure time.}
\tablenotetext{[2]}{Heliocentric distance.}
\tablenotetext{[3]}{Topocentric distance.}
\tablenotetext{[4]}{Phase angle.}
\tablenotetext{[5]}{Elongation.}
\tablenotetext{[6]}{Air mass, dimensionless. The range gives the values at the start and the end of the observation from each night, rather than the minimum and maximum.}
\tablenotetext{[7]}{C = cloud interference.}

\tablecomments{
All the observations exploited the Johnson {\it BVRI}~filters.
}
\end{deluxetable}

\clearpage

\begin{deluxetable}{ccccccc}
\tabletypesize{\footnotesize}
\tablecaption{Photometry of 2018 C2 from Xingming Observatory
\label{tab_phot}}
\tablewidth{0pt}
\tablehead{ 
\colhead{Date (UT)} & \colhead{$m_R \left(r_{\rm H}, {\it \Delta}, \alpha \right)$\tablenotemark{[1]}}
 & \colhead{$m_R \left(1,1,0 \right)$\tablenotemark{[2]}}  & \colhead{$m_B - m_R$}
 & \colhead{$m_B - m_V$}  & \colhead{$m_V - m_R$} & \colhead{$m_R - m_I$}
}
\startdata

2018 Mar 11 & $17.55 \pm 0.04$ & $13.88 \pm 0.04$ & $1.16 \pm 0.09$ & $0.71 \pm 0.09$ & $0.46 \pm 0.06$ & -- \\
2018 Mar 13 & $17.62 \pm 0.05$ & $13.99 \pm 0.05$ & $1.15 \pm 0.11$ & -- & -- & -- \\
2018 Mar 18 & $17.32 \pm 0.02$ & $13.77 \pm 0.02$ & $1.15 \pm 0.05$ & $0.74 \pm 0.05$ & $0.41 \pm 0.04$ & $0.36 \pm 0.04$ \\
2018 Mar 19 & $17.31 \pm 0.03$ & $13.77 \pm 0.03$ & $1.13 \pm 0.05$ & $0.74 \pm 0.05$ & $0.40 \pm 0.04$ & $0.38 \pm 0.05$ \\
2018 Mar 20 & $17.29 \pm 0.03$ & $13.77 \pm 0.03$ & $1.16 \pm 0.05$ & $0.77 \pm 0.05$ & $0.39 \pm 0.04$ & $0.38 \pm 0.06$ \\

\enddata

\tablenotetext{[1]}{Apparent {\it R}-band magnitude and $\pm 1\sigma$ uncertainty.}
\tablenotetext{[2]}{Absolute {\it R}-band magnitude.}

\tablecomments{
The {\it I}-band images from UT 2018 March 11 and 13, and {\it V}-band images from UT 2018 March 13 were discarded, because they were more influenced by clouds. Magnitude uncertainties are determined from photon statistics and error propagation.
}
\end{deluxetable}

\clearpage

\begin{deluxetable}{l|cccc}
\tabletypesize{\footnotesize}
\tablecaption{Orbital Elements of 2017 U7
(Heliocentric Ecliptic J2000.0)
\label{tab_orb_17U7}}
\tablewidth{0pt}
\tablehead{ Orbital Element & 
\multicolumn{2}{|c}{This Work}  & 
\multicolumn{2}{c}{JPL Horizons\tablenotemark{\dagger}} \\
 & 
\colhead{Value} & \colhead{1$\sigma$ Uncertainty} & 
\colhead{Value} & \colhead{1$\sigma$ Uncertainty} 
}
\startdata
Perihelion distance $q$ (AU) 
       & 6.418894 & 2.94$\times$10$^{-4}$ 
       & 6.418745 & 7.88$\times$10$^{-4}$ \\ 
Orbital eccentricity $e$ 
       & 1.001766 & 6.55$\times$10$^{-5}$ 
       & 1.001757 & 1.23$\times$10$^{-4}$ \\ 
Orbit inclination $i$ (\degr) 
       & 142.63894 & 4.42$\times$10$^{-5}$ 
       & 142.63896 & 1.56$\times$10$^{-4}$ \\ 
Longitude of ascending node $\Omega$ (\degr)
                 & 276.21700 & 4.85$\times$10$^{-4}$ 
                 & 276.21699 & 1.78$\times$10$^{-3}$ \\ 
Argument of perihelion $\omega$ (\degr)
                 & 325.94976 & 4.78$\times$10$^{-3}$ 
                 & 325.95210 & 1.20$\times$10$^{-2}$ \\ 
Time of perihelion $t_\mathrm{p}$ (TT)
                  & 2019 Sept 10.3996 & 4.09$\times$10$^{-2}$ 
                  & 2019 Sept 10.4187 & 9.70$\times$10$^{-2}$ \\ 

\enddata
\tablenotetext{\dagger}{The solution was retrieved on 2018 Mar 06.}
\tablecomments{
Both solutions have orbital elements at a common epoch of JD 2458086.5 = 2017 November 29.0 TT. The total numbers of astrometric observations, which span from UT 2017 August 18 to 2018 February 10, are 87 in both solutions. The weighted rms of our solution is $\pm0\arcsec.208$, whereas JPL Horizons only shows a dimensionless normalized rms of $\pm0.200$ for its solution.
}
\end{deluxetable}

\begin{deluxetable}{l|cccc}
\tabletypesize{\footnotesize}
\tablecaption{Orbital Elements of 2018 C2
(Heliocentric Ecliptic J2000.0)
\label{tab_orb_18C2}}
\tablewidth{0pt}
\tablehead{ Orbital Element & 
\multicolumn{2}{|c}{This Work}  & 
\multicolumn{2}{c}{JPL Horizons\tablenotemark{\dagger}} \\
 & 
\colhead{Value} & \colhead{1$\sigma$ Uncertainty} & 
\colhead{Value} & \colhead{1$\sigma$ Uncertainty} 
}
\startdata
Perihelion distance $q$ (AU) 
       & 1.956498 & 1.89$\times$10$^{-5}$ %
       & 1.956540 & 6.30$\times$10$^{-5}$ \\ %
Orbital eccentricity $e$ 
       & 1.001710 & 3.24$\times$10$^{-6}$ %
       & 1.001694 & 1.51$\times$10$^{-5}$ \\ %
Orbit inclination $i$ (\degr) 
       & 34.45301 & 2.76$\times$10$^{-5}$ %
       & 34.45301 & 1.86$\times$10$^{-4}$ \\ %
Longitude of ascending node $\Omega$ (\degr)
                 & 91.14150 & 8.75$\times$10$^{-5}$ %
                 & 91.14274 & 1.76$\times$10$^{-3}$ \\ %
Argument of perihelion $\omega$ (\degr)
                 & 134.14027 & 3.02$\times$10$^{-4}$ %
                 & 134.13622 & 4.85$\times$10$^{-3}$ \\ %
Time of perihelion $t_\mathrm{p}$ (TT)
                  & 2018 Jun 02.2314 & 4.07$\times$10$^{-4}$ %
                  & 2018 Jun 02.2268 & 4.82$\times$10$^{-3}$ \\ %

\enddata
\tablenotetext{\dagger}{The solution was retrieved on 2018 May 14.}
\tablecomments{
Both solutions have orbital elements at a common epoch of JD 2458164.5 = 2018 March 16.0 TT. The total number of astrometric observations in our solution is 525, and that in JPL Horizons' is 460. Both solutions have the earliest data from UT 2018 January 28. The latest observation in the solution by JPL Horizons was from UT 2018 April 26, while ours has an arc almost three weeks longer. The weighted rms of our solution is $\pm0\arcsec.059$, and JPL Horizons only shows a dimensionless normalized rms of $\pm0.272$ for its solution.
}
\end{deluxetable}

\end{document}